\makeatletter
\@ifundefined{@parse@version@dash}{%
\def\@parse@version#1{\@parse@version@0#1}
\def\@parse@version@#1/#2/#3#4#5\@nil{%
\@parse@version@dash#1-#2-#3#4\@nil}
\def\@parse@version@dash#1-#2-#3#4#5\@nil{%
  \if\relax#2\relax\else#1\fi#2#3#4 }
}{}
\makeatother

\documentclass[aps,prl,twocolumn,groupedaddress]{revtex4}

\usepackage{epsf,amsmath,amssymb,graphicx,scalefnt,rotating,enumitem,fancyhdr}
\usepackage[english]{babel}
\usepackage{multirow}
\usepackage[final]{showlabels}
\usepackage{slashed}
\usepackage{filemod}
\usepackage[pdftex, colorlinks=true, linkcolor=black, filecolor=black,
  citecolor=black, urlcolor=black, draft=false, bookmarks,
  bookmarksnumbered=true, plainpages=false,
  linktocpage={true}]{hyperref}
\usepackage[usenames,dvipsnames]{xcolor}

\newcounter{notecount}

\newcommand{\citere}[1]{Ref.\,\cite{#1}}
\newcommand{\citeres}[1]{Refs.\,\cite{#1}}

\newcommand{\abbrev}[1]{{\scalefont{.9}#1}}

\newcommand{\eqn}[1]{Eq.\,(\ref{#1})}

\newcommand{\order}[1]{{\cal O}(#1)}
\newcommand{\nklo}[1]{\abbrev{N$^{#1}$LO}}

\renewcommand{\Re}{{\rm Re}}

\newcommand{\mhiggs}{M_\text{H}}
\newcommand{\mtop}{M_\text{t}}

\newcommand{\myacrodef}[3]{\newcommand{#1}{\abbrev{#2}}}
\myacrodef{\heft}{HEFT}{Higgs effective field theory}
\myacrodef{\doublevirtual}{VV}{double virtual}
\myacrodef{\realvirtual}{RV}{real-virtual}
\myacrodef{\doublereal}{RR}{double real}
\myacrodef{\qcd}{QCD}{Quantum Chromo Dynamics}
\myacrodef{\lhc}{LHC}{Large Hadron Collider}
\myacrodef{\ho}{HO}{higher orders}
\myacrodef{\lo}{LO}{leading order}
\myacrodef{\nlo}{NLO}{next-to-leading order}
\myacrodef{\nnlo}{NNLO}{next-to-next-to-leading order}
\myacrodef{\llog}{LL}{leading logarithmic}
\myacrodef{\nll}{NLL}{next-to-leading logarithmic}
\myacrodef{\nnll}{NNLL}{next-to-next-to-leading logarithmic}
\myacrodef{\pdf}{PDF}{parton density function}
\myacrodef{\sm}{SM}{Standard Model}
\myacrodef{\bsm}{BSM}{beyond-the-\ac{SM}}
\myacrodef{\mssm}{MSSM}{Minimal Supersymmetric \ac{SM}}
\myacrodef{\susy}{SUSY}{Supersymmetry}
\myacrodef{\dreg}{DREG}{Dimensional Regularization}
\myacrodef{\dred}{DRED}{Dimensional Reduction}
\myacrodef{\emt}{EMT}{energy-momentum tensor}

\usepackage{array}

\begin{document}



\newcommand{\RHheaderline}{\textsf{TTK-21-17 / P3H-21-031 --- May 2021 }}
\fancypagestyle{firstpage}
{
  \renewcommand{\headrulewidth}{0pt}
  \fancyhead[R]{\RHheaderline}
}

\title{Exact top-quark mass dependence in hadronic Higgs production}


\author{M. Czakon, R.V. Harlander, J. Klappert, M. Niggetiedt}
\affiliation{Institute for Theoretical Particle Physics and Cosmology,\\
RWTH Aachen University, 52056 Aachen, Germany}


\date{\today}

\begin{abstract}
The impact of the finite top-quark mass on the inclusive Higgs
production cross section at higher perturbative orders has been an open
question for almost three decades. In this paper, we report on the
computation of this effect at \abbrev{NNLO} \abbrev{QCD}.  For the
purely gluonic channel, it amounts to +0.62\% relative to the result
obtained in the \abbrev{HEFT} approximation. The formally sub-leading
partonic channels over-compensate this shift, leading to an overall
effect of $-0.32\%$ at a $pp$ collider energy of 13\,TeV, and $-0.16\%$
at 8\,TeV.  This result eliminates one of the main theoretical
uncertainties to inclusive Higgs production cross section at
the \abbrev{LHC}.
\end{abstract}


\maketitle
\thispagestyle{firstpage}


\section{Introduction}\label{sec:intro}

Gluon fusion is the dominant production process for a Standard Model
(\sm) Higgs boson at the Large Hadron Collider (\lhc).  Obviously, this
makes a firm understanding of the pertinent cross section mandatory for
current and future precision Higgs physics.

A comprehensive analysis of the theoretical prediction for the inclusive
gluon fusion cross section has been performed
in \citere{deFlorian:2016spz}. It arrives at an overall theoretical
uncertainty of about $\pm 5$\%, arising from six different sources, each
of which contributes roughly 1\%~\footnote{This does not include the
uncertainty due to the \pdf{}s and the numerical value of $\alpha_s$,
which are independent of the theory progress for this particular
process.}. Thus, a significant reduction of the theoretical uncertainty
cannot be achieved by eliminating a single source, but entails efforts
on several of them.

Indeed, two of these sources have recently been addressed. The first one
was due to the fact that the next-to-next-to-next-to-leading order
(\nklo{3}) \qcd\ corrections were based on their expansion around
$z\equiv 1- \mhiggs^2/\hat s=0$\,\cite{Anastasiou:2015vya}, giving rise
to an uncertainty from the truncation of this expansion at finite order
in $z$. Meanwhile, however, the exact dependence on $z$ has become
available~\cite{Mistlberger:2018etf}.  The second one originated from
the use of a factorization formula for the mixed \qcd/electro-weak
effects\,\cite{Anastasiou:2008tj,Actis:2008ug}. Recent progress
indicates that this issue will be settled in the near
future~\cite{Bonetti:2017ovy,Anastasiou:2018adr,Bonetti:2018ukf,
Becchetti:2020wof}. On the other hand, two other sources of uncertainty,
namely the missing higher-order terms in both the partonic cross section
and the parton density functions (\pdf{}s), will require further
technological advances before one can expect significant improvements.

The two remaining sources of uncertainty identified
in \citere{deFlorian:2016spz} are related to quark mass effects. Both of
them originate from the fact that the gluon-fusion process is induced by
quark loops. The \nklo{n} \qcd\ corrections therefore involve an
$(n+1)$-loop calculation with at least two external mass scales (the
Higgs and the quark mass, and possibly other quark masses from
additional closed loops). While the next-to-leading order (\nlo)
result for arbitrary quark masses has been available for almost three
decades\,\cite{Graudenz:1992pv}, radiative corrections beyond this order
were mostly restricted to top-loop induced
terms\,\cite{Harlander:2002wh,Anastasiou:2002yz,Ravindran:2003um}, which
make up around 95\% of the total cross section in the \sm. Their exact
evaluation is the topic of the current paper. Progress in approximating
bottom- and charm-quark effects beyond \nlo\ has been made recently
in \citeres{Melnikov:2016emg,Braaten:2017ukc,Lindert:2017pky,
Caola:2018zye,Anastasiou:2020vkr}.

It turns out that the dominant effect of the top-loop induced terms can
be accounted for in the so-called Higgs effective field theory (\heft)
approximation, which is defined by multiplying the leading order (\lo)
cross section by the higher order (\ho) correction factor in the limit
of an infinite top-quark mass,
$\mtop\to\infty$\,\cite{Dawson:1990zj,Djouadi:1991tka}, which we take to
be defined in the on-shell scheme throughout this paper:
\begin{equation}\label{eq:kail}
  \begin{split}
    \sigma^\mathrm{HO}_\mathrm{HEFT} =
    \left(\frac{\sigma^{\mathrm{HO}}}{\sigma^\mathrm{LO}}\right)_{\mtop\to\infty}
    \sigma^\mathrm{LO} \; .
  \end{split}
\end{equation}
In this limit, the top-quark loop assumes the form of an effective
Higgs-gluon vertex\,\cite{Chetyrkin:1997un}, thus reducing the number of
associated loop integrations by one.  At \nlo, \eqn{eq:kail}
approximates the full hadronic cross section for a \sm\ Higgs boson to
about 0.1\%.  This is remarkable for several reasons.  On the one hand,
the assumption that $\mtop$ is the largest dimensional scale of the
process is invalid over a large range of the partonic center-of-mass
energy $\sqrt{\hat s}$, which reaches up to the collider energy
$\sqrt{s}\gg \mtop$. On the other hand, less than 50\% of the total
cross section is due to the \lo\ contribution, which means that the
$\mtop\to\infty$ approximation is applied to more than half of the total
cross section.  And finally, for the non-$gg$ partonic channels such as
$qg$ or $q\bar q$ the \heft\ approximation largely fails to capture the
top-mass effects at \nlo. It is only due to the strong numerical
dominance of the $gg$ channel that this hardly affects the total
hadronic cross section.

Qualitatively, the high accuracy of the \heft\ result can be explained
by the suppression of the large-$\hat s$ region by the \pdf{}s. Also a
dominance of the soft region in the total cross section could be made
responsible for the small impact of the top-mass effects at higher
orders. However, a solid quantitative understanding of this observation is
still missing.  The main reason for this is that higher order terms in
$1/\mtop$ introduce positive powers of $\hat s$ and thus
\textit{enhance} the large-$\hat s$ region. Consequently, such terms
cannot serve as uncertainty estimates of the heavy-top limit in a
straightforward way.

So far the only estimate of top-mass effects beyond the \heft\
approximation is therefore based on a combination of the
$1/\mtop$-expansion with the leading terms in the large-$\hat s$
limit\,\cite{Marzani:2008az,Harlander:2009my,Harlander:2009mq,Pak:2009dg},
from which an uncertainty of 1\% due to top-quark mass effects was
derived \cite{deFlorian:2016spz}.

In this paper we eliminate this uncertainty by reporting on the exact
calculation of the top-quark mass effects in hadronic Higgs production
at next-to-next-to-leading order (\nnlo) \qcd.


\section{Calculation}

The calculation requires the combination of the purely virtual
three-loop corrections to the cross section with the contributions from
the real emission both of a single parton (quark $q$, anti-quark $\bar
q$, or gluon $g$) at two-loop level, and of two partons at one-loop
level. Factorization scheme dependence demands to take all possible
partonic initial states into account. This is also important in the
light of the failure of \eqn{eq:kail} for the non-$gg$ channels as
mentioned before, combined with the fact that they increase by roughly
100\% from \nlo\ to \nnlo\ within \heft.

Calculations of all the relevant amplitudes, including their
full top-mass dependence, have already been reported on in the
literature. In fact, the double-real emission amplitudes have been
known for two decades~\cite{DelDuca:2001fn}. Today, they can be obtained
with public automated tools, and we use 
\texttt{OpenLoops} \cite{Buccioni:2019sur} for this purpose.

Complete results for the three-loop virtual amplitude are very recent.
Its full top-mass dependence at \nnlo\ has been first obtained with the
help of Pad\'e approximants constructed from the heavy-top expansion and
the non-analytic terms at the threshold $\hat
s=4\mtop^2$~\cite{Davies:2019nhm}. For the present study, we use a
numerical result that was derived by expressing the amplitude in terms
of master integrals, and subsequently evaluating them
numerically~\cite{Czakon:2020vql}.  Note that a fully analytic result is
only available for the part which involves light (massless) fermion
loops~\cite{Harlander:2019ioe}.

The main obstacle when calculating the total cross section with full
top-mass dependence are the two-loop single-emission
amplitudes. Unfortunately, existing results are not suitable for our
purpose. After a number of approximate
results\,\cite{Lindert:2018iug,Neumann:2018bsx}, the amplitudes have
been evaluated including their full top-mass dependence in the context
of the Higgs-plus-jet production\,\cite{Jones:2018hbb,Kerner:2019qgb},
but their numerical accuracy is insufficient, in particular since we
need them also in the soft and collinear regions. Semi-analytic results
for the master integrals in the form of one-dimensional generalised
power series have been presented as well~\cite{Frellesvig:2019byn}. In
lack of a public code for these results, it would be necessary to
implement the algorithm of \citere{Frellesvig:2019byn} from scratch,
which is a very demanding task.

In order to arrive at the required numerical precision, we have
calculated the single-emission contribution by following the strategy of
\citere{Czakon:2020vql}, which itself is based
on \citere{Czakon:2015exa}. In short, the amplitudes have been reduced
to a set of master integrals with the help of the public
software \texttt{Kira}$\oplus$\texttt{FireFly}\,\cite{Maierhoefer:2017hyi,
Maierhofer:2018gpa,Klappert:2019emp,Klappert:2020aqs,Klappert:2020nbg}.
Algebraic manipulations have been simplified by setting the ratio of the
top-quark and Higgs-boson mass to a fixed value of $\mtop^2/\mhiggs^2 =
23/12$, corresponding to $\mtop \approx 173.055$ GeV for $\mhiggs = 125$
GeV. The same software has also been used to derive a system of
first-order homogeneous linear differential equations in $\mtop$
satisfied by the master integrals. Using initial conditions in the
heavy-top limit, obtained with a diagrammatic large mass expansion, the
system of differential equations has been solved numerically at a very
large number of phase-space points. As a result, the amplitudes have
been obtained on a dense grid that could, in principle, be used for
interpolation. However, since the grid does not extend to the
boundaries of the phase-space where the amplitudes diverge, direct
inclusive phase-space integration requires a non-trivial extrapolation
to the singular soft and collinear regions.

\begin{figure}[t]
  \centering
  \includegraphics[width=.225\textwidth]{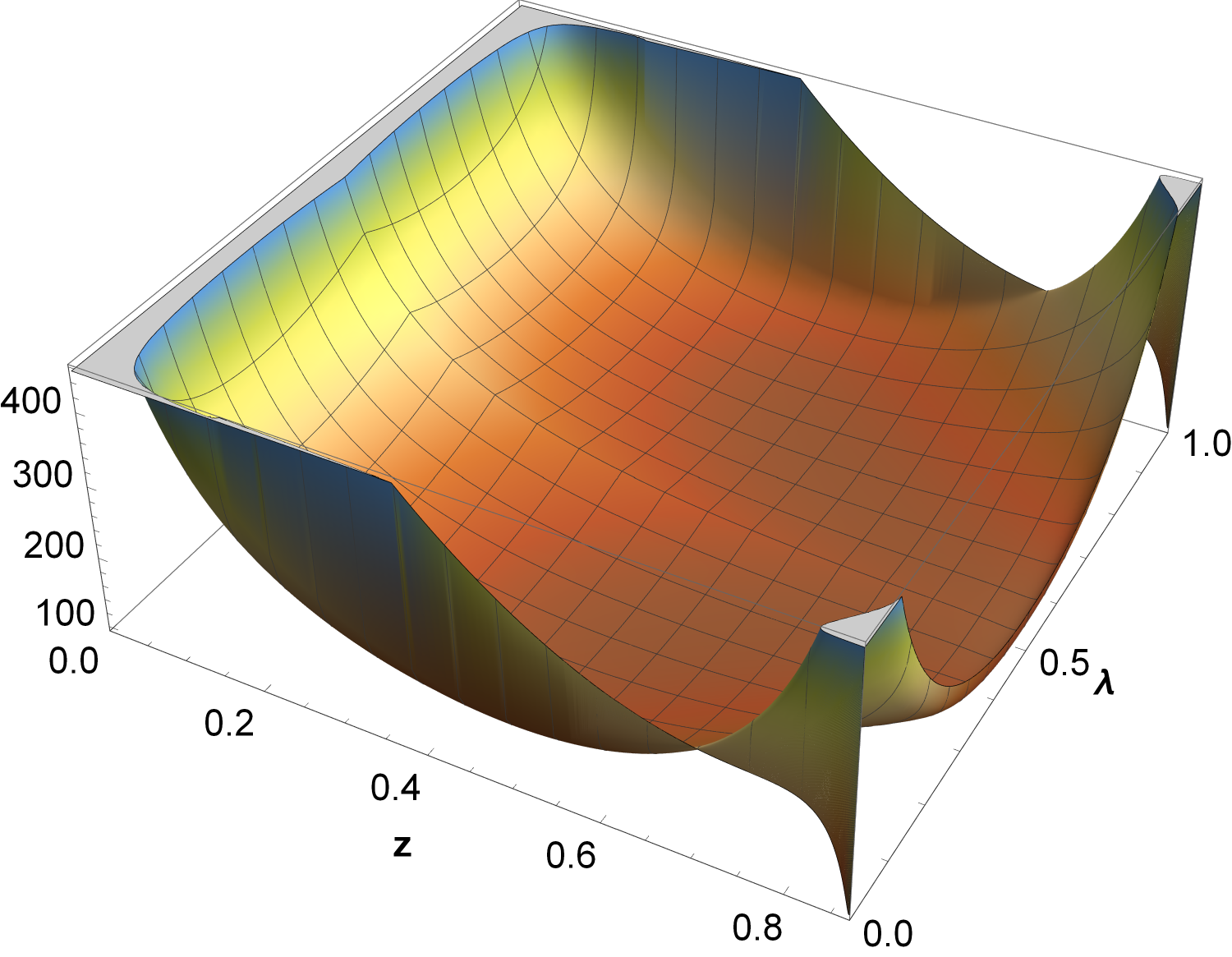}
  \includegraphics[width=.225\textwidth]{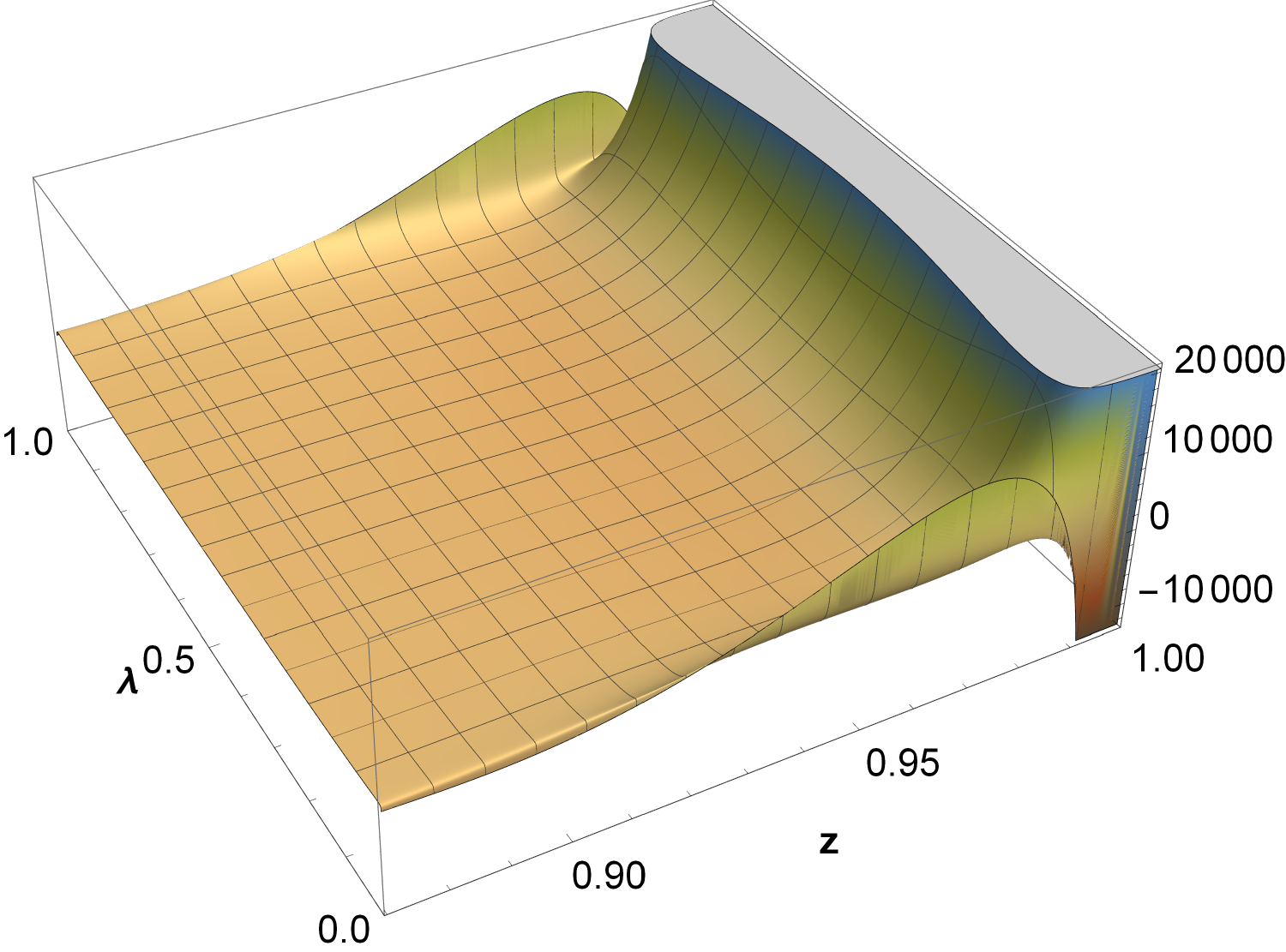}
  \includegraphics[width=.225\textwidth]{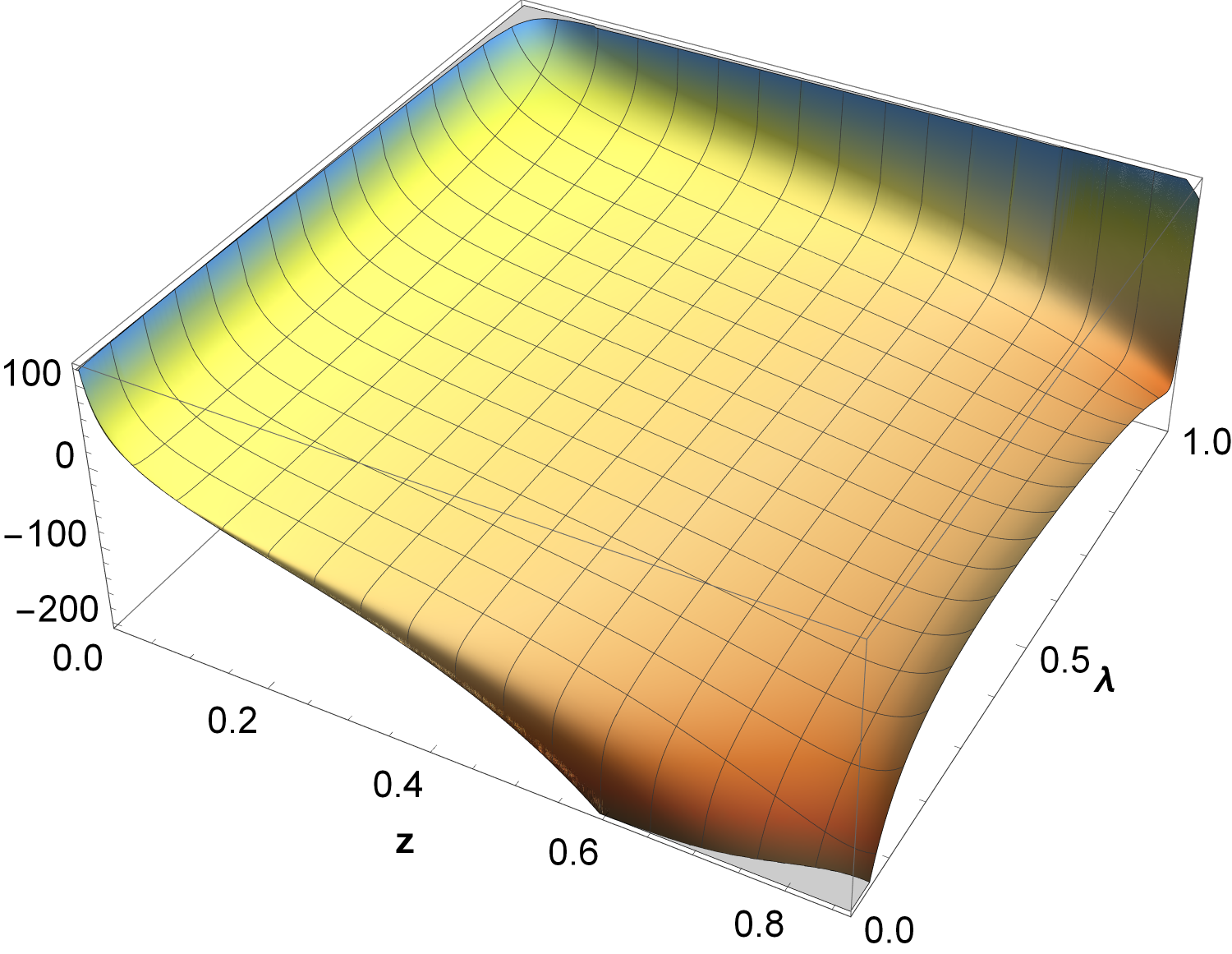}
  \includegraphics[width=.225\textwidth]{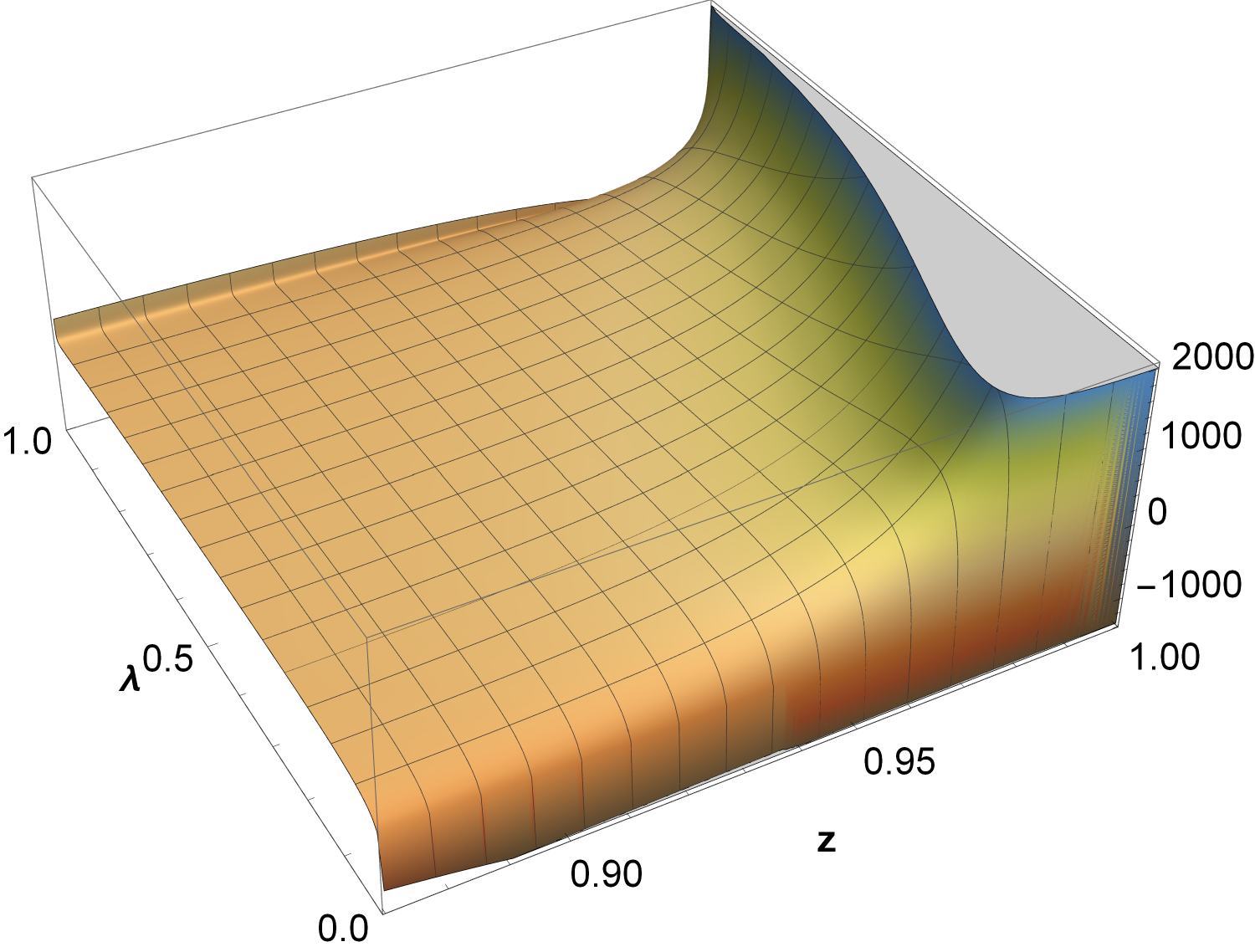}
  \includegraphics[width=.225\textwidth]{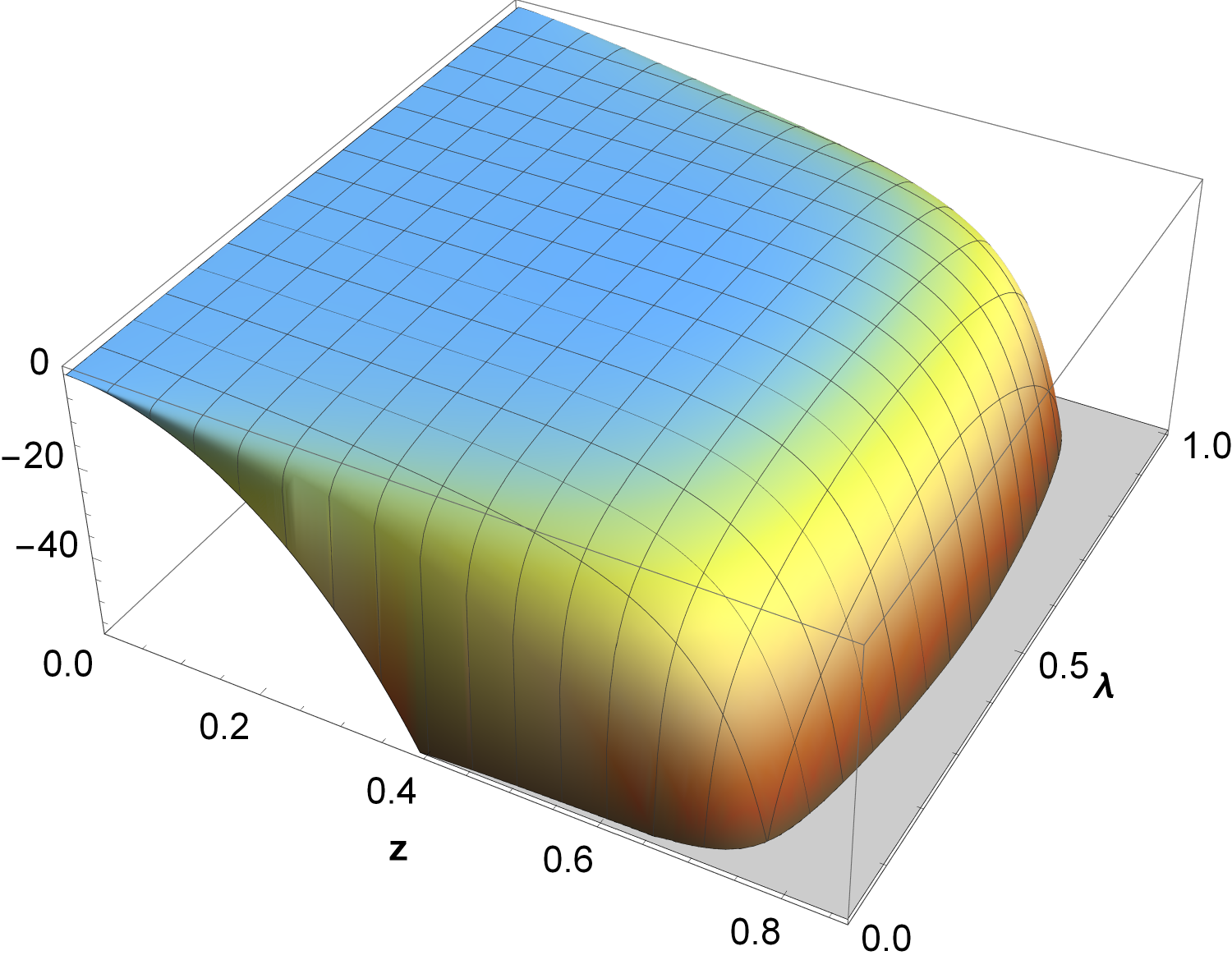}
  \includegraphics[width=.225\textwidth]{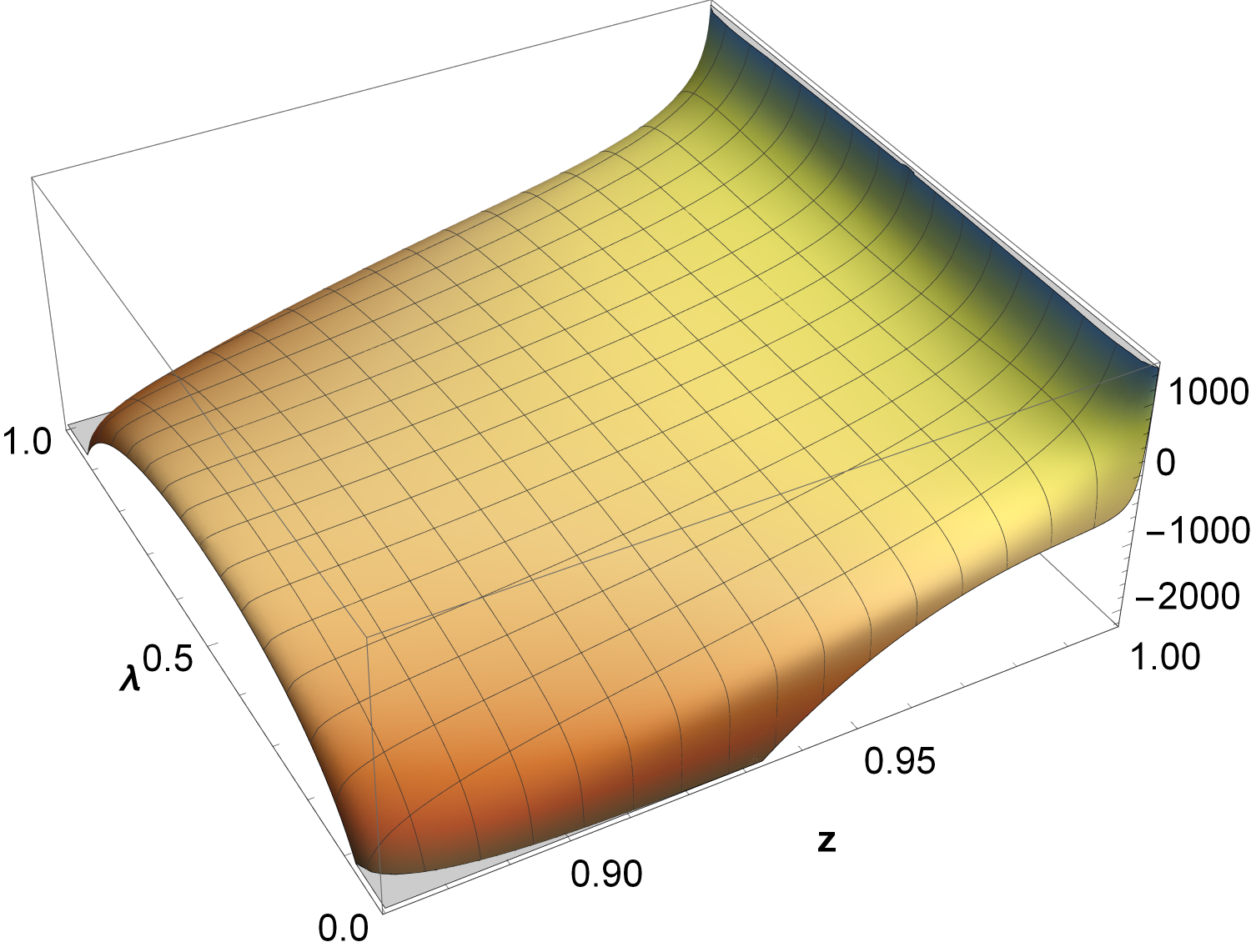}
  \caption{\label{fig:amplitudes} Finite part of the regulated
    amplitudes, $2 \Re \langle M^{(1)}_{\text{exact}} | M^{(2)}_{\text{exact}}
    \rangle \big|_{\text{regulated}}$, defined in
    \eqn{eq:trick}, for the processes $gg \to gH$
    (first row), $qg \to qH$ (second row) and $q\bar{q} \to gH$ (third
  row), separated into the region below (left column) and above (right column)
  threshold for intermediate top-quark pair production, $\hat s=4\mtop^2$. A factor of
  $\alpha_s^4/(4\pi)^2\cdot 1/v^2\cdot \hat{s}$, with $v$
  the Higgs-field vacuum expectation value, has been factored out. The
  renormalization scale has been set at $\mu_R = \mhiggs/2$. The
  kinematics is parameterized with $z \equiv 1-M_H^2/\hat{s}$ and $\lambda
  \equiv \hat{t}/(\hat{t}+\hat{u})$. $\hat{s}$, $\hat{t}$ and
  $\hat{u}$ are the standard Mandelstam variables.}
\end{figure}

In order to evaluate the phase-space integrals, the amplitudes have been
regulated in the soft and collinear limits by subtracting their
counterparts in \heft. For instance, for the $gg \to gH$ process
assuming $|\hat{t}| < |\hat{u}|$, with $\hat{t}, \hat{u}$ the partonic
Mandelstam variables, the contraction of the two-loop, $|
M^{(2)} \rangle$, and one-loop, $| M^{(1)} \rangle$, amplitudes, both
treated as vectors in color and spin space, has been replaced with:
\begin{widetext}
\begin{equation} \label{eq:trick}
  \langle M^{(1)}_{\text{exact}} | M^{(2)}_{\text{exact}} \rangle \big|_{\text{regulated}} \equiv
  \langle M^{(1)}_{\text{exact}} |M^{(2)}_{\text{exact}}  \rangle -
  \Bigg[\langle M^{(1)}_{\text{HEFT}} | M^{(2)}_{\text{HEFT}} \rangle
  + \frac{8\pi \alpha_s}{\hat{t}} \Big\langle
  P_{gg}^{(0)}\Big(\frac{\hat{s}}{\hat{s} + \hat{u}} \Big) \Big\rangle
  \langle F^{(1)} | \big( F^{(2)}_{\text{exact}}  -
  F^{(2)}_{\text{HEFT}} \big) \rangle \Bigg] \; ,
\end{equation}
\end{widetext}
with $|F^{1,2} \rangle$ the respective amplitudes for the $gg \to
H$ process, and $\langle P_{gg}^{(0)}(z) \rangle$ the spin-averaged
Altarelli-Parisi splitting function:
\begin{equation}
  \langle P_{gg}^{(0)}(z) \rangle = 2C_A \Big( \frac{z}{1-z} + \frac{1-z}{z} + z(1-z) \Big) \; .
\end{equation}
It is not difficult to convince oneself that $\langle
M^{(1)}_{\text{exact}} | M^{(2)}_{\text{exact}} \rangle \big|_{\text{regulated}}$ is
devoid of soft and collinear singularities. We use an analogous expression for the $qg \to qH$
process with:
\begin{equation}
  -\langle P_{qq}^{(0)}(z) \rangle =-T_F \big( 1-2z(1-z) \big) \; ,
\end{equation}
instead of $\langle P_{gg}^{(0)}(z) \rangle$, while there is no
splitting-function contribution in the $q\bar{q} \to gH$ case. The
regulated amplitudes for the three processes are illustrated in
Fig.~\ref{fig:amplitudes}. Notice that the amplitudes for the first two
processes are still singular in the collinear limit, but these
singularities are integrable and occur very close to the edge of the
phase space. In order to obtain a reliable inclusive phase-space
integral, we approximate the regulated amplitudes in the limit
$\hat{t} \to 0$ with the ansatz $a \ln |{\hat{t}}| + b$, for each value
of $z$, and fit the coefficients to the available numerical values of
the amplitudes at $|{\hat{t}}| > |{\hat{t}_0}|$. The ansatz is
subsequently integrated analytically in the region $0 > |{\hat{t}}| >
|{\hat{t}_0}|$. The uncertainty of the procedure is estimated by using
the more general ansatz, $a \ln |{\hat{t}}| + b + c \hat{t} \ln
|{\hat{t}}| + d \hat{t}$.

Having the amplitudes for all contributions at hand, it is necessary
to actually integrate them in order to obtain the cross section
contributions. Since the effect of the top-quark mass beyond the
heavy-top limit is expected to be small, we directly evaluate the
difference of the cross sections at each phase-space point:
\begin{equation} \label{eq:diff}
\int \left(\mathrm{d} \sigma_{\text{exact}}^{\text{(N)NLO}} - \mathrm{d} \sigma_{\text{HEFT}}^{\text{(N)NLO}}\right)
\; ,
\end{equation}
rather than the cross sections themselves separately. This has the
additional advantage that ultraviolet and infrared divergences in the
form of $1/\epsilon^k$ poles in the dimensional regularization
parameter, $\epsilon$, as well
as soft and collinear singularities first appear at the \nnlo\
level. Hence, for example, the \nlo\ contributions to the difference are
well-defined separately for the virtual and real corrections. This delay
of the appearance of divergences and singularities is one of the reasons
for the smallness of the top-quark mass effects beyond \heft.

Ultimately, \eqn{eq:diff} is evaluated with Monte Carlo methods
using the sector-improved residue subtraction scheme
\cite{Czakon:2010td,Czakon:2014oma,Czakon:2019tmo} implementation in
the \texttt{C++} library \texttt{Stripper}. Note that it suffices to use
the subtraction term in the square brackets of \eqn{eq:trick} in order
to cancel the \abbrev{IR} divergences with the double real
emission. Since this subtraction term is given in terms of compact analytic
formulae~\cite{Schmidt:1997wr}, it allows for a fast and numerically
stable Monte Carlo integration. The phase space integration and \pdf{}
convolution of the r.h.s.\ of \eqn{eq:trick} is done separately.  Adding
it to the output from \texttt{Stripper} cancels the subtraction term
contribution and leads to the final result.


\section{Results}

\newcolumntype{x}[1]{>{\raggedleft\hspace{0pt}}p{#1}}
\begin{table*}[t!]
  \caption{\label{tab:exact-heft}Effects of a finite top-quark mass on
  the total hadronic Higgs-boson production cross section for the \lhc{}
  @ 13~TeV and 8~TeV, separately for the partonic channels and including
  Monte Carlo integration error estimates. Results obtained with
  the \pdf\ set \texttt{NNPDF31\_nnlo\_as\_0118}~\cite{Ball:2017nwa},
  renormalization and factorization scales $\mu_R = \mu_F = \mhiggs/2$,
  Higgs-boson mass $\mhiggs = 125$ GeV, and top-quark mass $\mtop
  = \sqrt{23/12} \times \mhiggs \approx 173.055$ GeV. The \nnlo\ cross
  section within \heft\ ($\sigma^\textrm{NNLO}_\mathrm{HEFT}$) has been
  obtained with \texttt{SusHi}~\cite{Harlander:2012pb,Harlander:2016hcx}
  and is split into contributions from the individual orders in
  $\alpha_s$.}
\begin{tabular}{c|x{.2\textwidth}|x{.14\textwidth}x{.2\textwidth}|>{\centering\arraybackslash}p{.2\textwidth}}
\hline
\rule{0pt}{1em}
\multirow{2}{*}{channel} & \multicolumn{1}{c|}{$\sigma^\mathrm{NNLO}_\mathrm{HEFT}$ [pb]} & 
\multicolumn{2}{c|}{$(\sigma^\mathrm{NNLO}_\mathrm{exact}-
\sigma^\mathrm{NNLO}_\mathrm{HEFT})$ [pb]} &
\multirow{2}{*}{$(\sigma^\mathrm{NNLO}_\mathrm{exact} / \sigma^\mathrm{NNLO}_\mathrm{HEFT} -1)$ [\%]} \\
& $\order{\alpha_s^2}+\order{\alpha_s^3}+\order{\alpha_s^4}$& 
\multicolumn{1}{r}{$\order{\alpha_s^3}$} &
\multicolumn{1}{c|}{$\order{\alpha_s^4}$} &
\\
\hline
\multicolumn{5}{c}{\rule{0pt}{1em}$\sqrt{s} = 8$\,TeV}\\\hline
$gg$ & $7.39 + 8.58 + 3.88$ &  $+0.0353$ & $+0.0879\pm 0.0005$ & $+0.62$ \\
$qg$ & $0.55 + 0.26$ &  $-0.1397$ & $-0.0153\pm 0.0002$ & $-19$ \\
$qq$ & $0.01 + 0.04$ &  $+0.0171$ & $-0.0191\pm 0.0002$ & $-4$ \\\hline
total & $7.39 + 9.14 + 4.18$ &  $-0.0873$ & $+0.0535\pm 0.0006$ & $-0.16$ \\
\hline
\multicolumn{5}{c}{\rule{0pt}{1em}$\sqrt{s} = 13$\,TeV}\\\hline
$gg$ & $16.30 + 19.64 + 8.76$ &  $+0.0345$ & $+0.2431\pm 0.0020$ & $+0.62$ \\
$qg$ & $1.49 + 0.84$ &  $-0.3696$ & $-0.0408\pm 0.0005$ & $-18$ \\
$qq$ & $0.02 + 0.10$ &  $+0.0322$ & $-0.0501\pm 0.0006$ & $-15$ \\\hline
total & $16.30 + 21.15 + 9.70$ &  $-0.3029$ & $+0.1522\pm 0.0021$ & $-0.32$ \\
\hline\end{tabular}
\end{table*}

Table~\ref{tab:exact-heft} collects our main results. It shows the
hadronic cross section $\sigma_\textrm{HEFT}^\textrm{NNLO}$ in
the \heft\ approximation through \nnlo\ \qcd, including only top-loop
induced contributions and without electro-weak effects, and separately
for the partonic sub-channels ($qq$ denotes the sum over all quark
initial states). The absolute numbers are split into the contributions
from the individual orders in $\alpha_s$. The uncertainties indicate the
Monte Carlo integration errors.

While the finite-mass effects are small and positive for the $gg$
channel (and largely independent of the collider energy), the relative
effect on the other channels is negative and much larger. For the pure
quark channels, the \heft\ approximation is off by more than 100\% at
 each perturbative order. Taken individually, this would already
exhaust the uncertainty estimate associated with the missing mass
effects of \citere{deFlorian:2016spz}, despite the fact that these
channels contribute to the total cross section only at the 1-2\%
level. In fact, we find that the absolute values of all finite-mass
effects add up to about 1.5-1.6\% at \nnlo. However, the cancellations among
the individual channels and perturbative orders decrease this number to
$-0.16$\% at 8\,TeV, and $-0.32$\% at 13\,TeV.


\section{Conclusions and Outlook}

A calculation of the hadronic Higgs production cross section including
the full top-mass dependence at \nnlo\ was reported. It results in a
slight decrease relative to the result in the \heft\ approximation of
$-0.32\%$ at 13\,TeV, and $-0.16\%$ at 8\,TeV collider energy. This
result confirms and at the same time eliminates the commonly accepted
uncertainty estimate arising from the lack of knowledge of these
effects.

Our calculational techniques are also applicable to the bottom- and
charm-loop induced terms and the associated interference
with the top-loop terms. This is deferred to future work.

\begin{acknowledgments}
This research was supported by the Deutsche Forschungsgemeinschaft (DFG,
German Research Foundation) under grants 396021762 - TRR 257 and 400140256 - GRK 2497: The physics of the heaviest particles at the Large Hardon Collider.

Simulations were performed with computing resources granted by RWTH
Aachen University under projects rwth0414 and rwth0643.
\end{acknowledgments}


\bibliography{ggh}

\end{document}